\documentclass[aps,preprintnumbers,amsmath,amssymb,11pt]{revtex4}

\usepackage{epsf}

\usepackage{graphicx}

\usepackage{bm}

\begin{document}

%%%%%%%%%%%%%%%%%%%%%%%%%%%%%%%%%%%%%%%%%%%%%%%%%%%%%%%%%%%%%%%%%%%%%%%%%%%%%%%

\def\Qs{Q_{\rm s}}

\def\half{{\textstyle{\frac12}}}
\def\p{{\bm p}}
\def\q{{\bm q}}
\def\x{{\bm x}}\def\v{{\bm v}}
\def\E{{\bm E}}
\def\B{{\bm B}}
\def\A{{\bm A}}
\def\a{{\bm a}}
\def\b{{\bm b}}
\def\c{{\bm c}}
\def\j{{\bm j}}
\def\n{{\bm n}}
\def\grad{{\bm\nabla}}
\def\da{d_{\rm A}}
\def\tr{\operatorname{tr}}
\def\Im{\operatorname{Im}}
\def\md{m_{\rm D}}
\def\mpl{m_{\rm pl}}
\def\pol{\varepsilon}
\def\bpol{{\bm\pol}}
\def\CP{CP^{N-1}}
\def\be{\begin{equation}}
\def\ee{\end{equation}}
\def\bea{\begin{eqnarray*}}
\def\eea{\end{eqnarray*}}

%%%%%%%%%%%%%%%%%%%%%%%%%%%%%%%%%%%%%%%%%%%%%%%%%%%%%%%%%%%%%%%%%%%%%%%%%%%%%%%

\title
    {
Anomaly Inflow and Membrane Dynamics in the QCD Vacuum
    }

\author {H. B. Thacker\footnotemark \footnotetext{email: hbt8r@virginia.edu}$~^a~$ and Chi Xiong\footnotemark \footnotetext{email: xiongchi@ntu.edu.sg}$~^b$
}
\affiliation
    {%
 {\bf a.}~Department of Physics,
    University of Virginia,
    P.O. Box 400714
    Charlottesville, VA 22901-4714, USA\\
  {\bf b.}~Institute of Advanced Studies, Nanyang Technological University, \\
     Singapore 639673  
}

%\date{\today}

\begin {abstract}%
    {%
Large $N_c$ and holographic arguments, as well as Monte Carlo results, suggest that the topological
structure of the QCD vacuum is dominated by codimension-one membranes which appear as thin dipole layers of 
topological charge. Such membranes arise naturally as $D6$ branes in the holographic formulation of QCD based on IIA string theory. 
The polarizability of these membranes leads to a vacuum energy $\propto \theta^2$, providing the origin
of nonzero topological susceptibility. Here we show that the axial $U(1)$ anomaly can be formulated as anomaly inflow
on the brane surfaces. A 4D gauge transformation at the brane surface separates into a 3D gauge
transformation of components within the brane and the transformation of the transverse component. The in-brane gauge transformation 
induces currents of an effective Chern-Simons theory on the brane surface,
while the transformation of the transverse component describes the transverse motion of the brane and is related to the
Ramond-Ramond closed string field in the holographic formulation of QCD. The relation between the surface currents and the transverse
motion of the brane is dictated by the descent equations of Yang-Mills theory.
}% 
\end {abstract}

\maketitle
\thispagestyle {empty}

 %%%%%%%%%%%%%%%%%%%%%%%%%%%%%%%%%%%%%%%%%%%%%%%%%%%%%%%%%%%%%%%%%%%%%%%%%%%%%%%

\section {Introduction}

The possible importance of codimension one membrane-like topological charge structures in the QCD vacuum is suggested by both 
theoretical considerations \cite{Witten79,Witten98} and by Monte Carlo studies \cite{Horvath03,Ilgenfritz}. Theoretically, the suggestion of topological
domain wall structures in the vacuum emerged from large-$N_c$ chiral Lagrangian arguments. These arguments showed that, in the 
large-$N_c$ limit the multivaluedness of the effective $\eta'$ mass term induced by the chiral $U(1)$ anomaly implies the existence of multiple, discrete, quasistable, and nearly
degenerate ``k-vacua''. These vacua are labeled by effective local values of the QCD $\theta$ parameter which differ by integer multiples of $2\pi$,
and are separated by domain walls where the value of $\theta$ jumps by $\pm 2\pi$.
With the emergence of the holographic string theory framework for QCD-like gauge theories \cite{Maldacena,Polyakov,Witten98}, it was shown that the role of the domain wall predicted by
large $N_c$ was played by the $D6$ brane of IIA string theory. (More precisely, by an ``$I2$ brane'' which is the intersection of the $D6$ brane
with the $D4$ color branes \cite{Witten98}.) In the holographic framework the local $\theta$ parameter is given by the Wilson line of the closed string Ramond-Ramond
U(1) gauge field around the compactified direction of the $D4$ branes. In 9+1 dimensions, a $D6$ brane plays the role of a magnetic monopole source for the RR field, 
and is dual to an instanton, which is represented by a $D0$ brane in the holographic model. The incompatibility of the instanton model with large $N_c$ chiral dynamics \cite{Witten79}
indicated that, at least for sufficiently large $N_c$, instantons should be replaced by codimension one membranes or domain walls.. 
From the 9+1-dimensional string viewpoint, the membrane-dominated vacuum is in a precise sense dual to the instanton vacuum, with
instantons (D0 branes) and domain walls (D6 branes) being, respectively, electric and magnetic sources of Ramond-Ramond field. 
The success of large $N_c$ phenomenology, combined with the Monte Carlo evidence for topological charge membranes in $SU(3)$ gauge theory \cite{Horvath03,Ilgenfritz}
strongly indicates that $N_c =3$ is large enough that the membrane-dominated vacuum is the correct qualitative picture for real QCD.
In 4-dimensional spacetime, a $D6$ brane appears as a 2+1 dimensional intersection with the color branes, with the other 4 spatial dimensions of the $D6$ brane 
compactified on an $S_4$. As a color excitation, the $D6$ brane appears as a codimension one dipole layer of topological charge in the gauge field. 
The quantized jump in the value of $\theta$ across the membrane is just the Dirac quantization condition for RR monopoles.
In this paper, we show that the existence and dynamics of these topological membranes in the QCD vacuum can be studied without reference to the higher-dimensional string theory framework,
using the descent equations and cohomology structure of 4-dimensional Yang-Mills theory \cite{Zumino83, Stora83, Faddeev84}. This provides a ``bottom-up'' perspective on the holographic framework.
It also identifies the exact mechanism by which the RR U(1) gauge field remains in 4-dimensional QCD as an auxiliary field representing singular membrane-like excitations of the Yang-Mills field.

We consider 4D SU(N) Yang-Mills theory and, in order to construct a domain wall, we add an external source coupled to topological charge,
\begin{equation}
S = S_{YM} + \int d^4x\; \theta(x)Q(x)
\end{equation}
where 
\begin{equation}
Q(x) =\frac{1}{16\pi^2N_c} \varepsilon^{\mu\nu\sigma\tau}Tr F_{\mu\nu}F_{\sigma\tau}
\end{equation}
is the Yang-Mills topological charge density, and $\theta(x)$ represents the local value of the theta parameter. 
We construct a straight flat membrane or domain wall by taking $\theta(x)$
to be a step function along one spatial direction, which we label $x_1$, with discontinuity $\theta_0$,
\begin{eqnarray}
\label{eq:theta}
\theta(x) & = & \theta_0 \;\;\; x_1>0 \\
        & = & 0  \;\;\;\; x_1<0
\end{eqnarray}
This produces a codimension one membrane occupying the dimensions transverse to $x_1$. Integrating by parts, we can write the source term as an integral on the surface of the brane,
where the integrand is proportional to the Chern-Simons current,
\begin{equation}
\label{eq:CScurrent}
K_{\mu} = \varepsilon_{\mu\alpha\beta\gamma}{\rm Tr}\left(A^{\alpha}\partial^{\beta}A^{\gamma} + \frac{2}{3}A^{\alpha}A^{\beta}A^{\gamma}\right)
\equiv \varepsilon_{\mu\alpha\beta\gamma}{\cal K}_3^{\alpha\beta\gamma}
\end{equation}
whose divergence is proportional to the topological charge density, 
\begin{equation}
\label{eq:anomaly2}
\partial^{\mu}K_{\mu}={32\pi^2N_c}Q(x)
\end{equation}
The term in the action is thus expressed in terms of Yang-Mills fields localized to the brane surface, but it is no longer explicitly gauge invariant (because we have discarded
a surface term at $x_1=\infty$). The gauge 
variation of the CS 3-form in (\ref{eq:CScurrent}) is dictated by the descent equations \cite{Zumino83, Stora83, Faddeev84}, 
and can be written formally as the exterior derivative of a 2-form. However, one of the terms in the resulting
2-dimensional surface integral has a topological ambiguity. 
It has the form of a Wess-Zumino-Witten term and is only defined modulo $2\pi$ times an integer which identifies the winding number of the gauge transformation in the
3-volume enclosed by the 2D surface. Gauge invariance of the exponentiated WZW term requires the integer quantization of the coupling constant, which in this case is $\theta_0/2\pi$.  
In this way we find that the quantization of the step in $\theta$ across the domain wall follows from the requirement of invariance under ``large'' gauge transformations, in a manner similar to 
the quantization of the WZW coupling constant in 2D sigma models. 

Next we show that by considering small, nontopological gauge transformations we obtain information
about the dynamics of fluctuating branes. This occurs because invariance under a 4D Yang-Mills gauge transformation provides a 
relation between the gauge variation of the Chern-Simons 3-form on the brane surface and the
transformation of the gauge field component transverse to the brane surface. The Chern-Simons form that appears in the surface integral at $x_1=0$ depends only on the three ``in brane'' components. The transverse
component of $A_{\mu}$ is related to membrane fluctuations in the transverse coordinate. A central role in this discussion is played by the descent equations of
4D Yang-Mills theory \cite{Zumino83, Stora83, Faddeev84}, which describes the intertwining of gauge cohomology with spacetime de Rahm cohomology by relating gauge variations to exterior derivatives.
For a curved or fluctuating membrane $\partial^{\mu}\theta$ is a vector normal to the surface. This vector thus specifies the local orientation of the brane surface. 
In order to construct gauge invariant amplitudes in the presence of a fluctuating brane, we define the auxiliary field $\partial^{\mu}\theta$ to transform under a Yang-Mills gauge transformation  
to cancel the variation of the Chern-Simons current on the brane surface:
\begin{equation}
\label{eq:delta_theta}
\delta(\partial_{\mu}\theta) = -\delta K_{\mu}
\end{equation}
The gauge invariance constraint (\ref{eq:delta_theta}) is an ``anomaly inflow'' condition that specifies the gauge variation of the RR field (i.e. of $\theta(x)$)
that must accompany a 4D Yang Mills gauge transformation. In a string theory framework, the idea of anomaly inflow is important for understanding the relation between string theory in the bulk, where only
closed strings propagate, and open-string gauge theories defined on lower dimensional brane surfaces. Anomalous, fermionic currents of the gauge theory are seen as currents which are
conserved overall, but which can flow onto and off of the brane surface, so only the combination of brane and bulk current is conserved. 
From the reverse perspective of the bulk theory, anomaly inflow is a generalization of the Dirac monopole
construction, with the Bianchi identity being preserved by a cancellation between the bulk magnetic flux and that carried away by the Dirac string. Normally one would expect
that the axial $U(1)$ anomaly in QCD could only be interpreted in terms of anomaly inflow by embedding it in a higher dimensional ``bulk'' theory, e.g. type IIA string theory in 10 dimensions.
However, the role of codimension one membranes in the vacuum of 4-dimensional QCD allows the anomaly inflow mechanism to be operative in a strictly 4-dimensional context,
with the spatial direction transverse to the membrane playing the role of a bulk coordinate.

In the case of the axial $U(1)$ anomaly in QCD the physical gauge invariant flavor singlet 
axial current $\hat{j}_5^{\mu}$ is not conserved, $\partial_{\mu}\hat{j}_5^{\mu}= Q(x)$, but it is sometimes convenient to introduce a conserved, non-gauge invariant
axial current $j_5^{\mu}$ by subtracting off the Chern-Simons current of the gauge field,
\begin{equation}
\label{eq:anomaly}
j_5^{\mu} = \hat{j}_5^{\mu} - \frac{1}{32\pi^2N_c}K^{\mu}
\end{equation}
which is conserved by Eq. (\ref{eq:anomaly2}).
Here we interpret this construction as an anomaly inflow constraint. If only the in-brane components of the gauge field are nonzero, the Chern-Simons current $K^{\mu}$ is
a vector transverse to the brane whose support is localized on the membrane surface. 
Away from the brane surface, in the 4-dimensional bulk, the axial current is both conserved and gauge invariant. The nonconservation of the current $\hat{j}_5^{\mu}$ 
occurs only on the brane surface, where it can be carried away in the form of surface currents associated with the gauge variation of the Chern-Simons 3-form.
 
As shown in Ref. \cite{TXK}, the anomaly inflow condition (\ref{eq:delta_theta}) enforces a Kogut-Susskind cancellation \cite{Kogut75} 
between massless poles coupled to the (separately non gauge invariant) operators
$\partial_{\mu}\theta$ and $K_{\mu}$, so that there are no massless poles coupled to the gauge invariant combination $\partial_{\mu}\theta + K_{\mu}$. 
In the holographic description, this invokes the anomaly inflow requirement that the gauge variation of the CS 3-form on the I2 brane should be cancelled by 
a gauge variation of the bulk RR field on the brane surface \cite{Green,TXK}. The Kogut-Susskind cancellation of massless poles in gauge invariant amplitudes 
is a manifestation of the gauge invariance that connects the in-brane components of the gauge field to the transverse component. As we will discuss in detail for the 2-dimensional case considered in 
Section II, this relates the Chern-Simons current on the brane to the local spacetime orientation of the brane surface. 

In 4D Yang-Mills theory, the gauge cancellation (\ref{eq:delta_theta}) specifies a relation between a 3D Yang-Mills transformation within the brane and a change of the local orientation of the 
brane surface, as determined by $\partial_{\mu}\theta$. In this way, we relate the variation of $\partial^{\mu}\theta$ to the 1-cocycle of the $SU(N)$ gauge transformation 
$g\equiv e^{i\omega}$, as constructed from the descent equations \cite{Zumino83, Stora83, Faddeev84}. For our purposes, a 1-cocycle can be thought of as a ``Berry phase'' associated
with transporting a representation of the gauge group around a closed orbit in the parameter space of gauge transformations.

The cocycle is a functional of the gauge transformation and is given by the integral of a local 2-form density over the surface of the brane at fixed time..
The 1-cocycle that is attached to a 2-dimensional brane surface is constructed from the topological charge by the descent procedure \cite{Faddeev84}.
The gauge variation of the Chern-Simons current is the sum of two terms,
\begin{equation}
\label{eq:CSvariation}
\delta K_{\mu} = \varepsilon_{\mu\alpha\beta\gamma}\left[\delta{\cal K}_{3A}^{\alpha\beta\gamma} + \delta{\cal K}_{3B}^{\alpha\beta\gamma}\right]
\end{equation}
where one term is the Maurer-Cartan form
\begin{equation}
\label{eq:K_3A}
\delta{\cal K}_{3A}^{\alpha\beta\gamma} = \frac{1}{3}{\rm Tr}\left[ g^{-1}\partial^{\alpha}g g^{-1}\partial^{\beta}g g^{-1}\partial^{\gamma}g \right]
\end{equation}
Since topological charge is gauge invariant, 
\begin{equation}
\delta(\partial^{\mu}K_{\mu}) = \partial^{\mu}\delta K_{\mu} = 0
\end{equation}
we expect that, locally, we can write the 3-form in (\ref{eq:CSvariation}) as an exterior derivative,
\begin{equation}
\label{eq:CSvariation2}
\varepsilon_{\mu\alpha\beta\gamma}\left[\delta{\cal K}_{3A}^{\alpha\beta\gamma} + \delta{\cal K}_{3B}^{\alpha\beta\gamma}\right]=
\varepsilon_{\mu\alpha\beta\gamma}\partial^{\alpha}\left[\delta{\cal K}_{2A}^{\beta\gamma} + \delta{\cal K}_{2B}^{\beta\gamma}\right]
\end{equation}
The Maurer-Cartan term can be written formally as the exterior derivative of a 2-form.
However, ${\cal K}_{2A}$ is not single-valued because it depends on the multivalued gauge phase $\omega=-i\ln\,g$. Up to terms of order $(\omega)^4$, 
it is given by
\begin{equation}
\label{eq:K_2A}
\delta{\cal K}_{2A}^{\beta\gamma} = \frac{i}{3}{\rm Tr}\left[ \omega g^{-1}\partial^{\beta}g g^{-1}\partial^{\gamma}g \right] + {\cal O}(\omega^4)
\end{equation}
As in the 2D WZW sigma model, the integral of this term over a closed 2D surface is ambiguous mod $2\pi$.
It depends not only on the values of $g$ on the 2-dimensional boundary, but on its winding number in the enclosed 3-dimensional volume. 
The second term in (\ref{eq:CSvariation}) describes an interaction between the WZW current and the Yang-Mills gauge potential,
\begin{equation}
\label{eq:K_2B}
\delta{\cal K}_{2B}^{\beta\gamma} = {\rm Tr}\left[ \partial^{\beta}g\;g^{-1}A^{\gamma} \right]
\end{equation}
This is a single-valued, nontopological contribution to the 1-cocycle. It describes the emission of gluons which accompanies brane fluctuations.

\section{Wilson lines as membranes in 2-dimensional U(1) gauge theory}

The basic idea of our formulation of brane dynamics in gauge theory is illustrated in a particularly simple context by the case of 2-dimensional $U(1)$ gauge theory. 
For this case, the Chern-Simons current is $K_{\mu}= \varepsilon_{\mu\nu}A^{\nu}$, and the analog of the descent equation (\ref{eq:CSvariation}) is simply related to the
gauge transformation $g=e^{i\omega}$ itself,
\begin{equation}
\delta K_{\mu} = -i\varepsilon_{\mu\nu}g^{-1}\partial^{\nu}g = \varepsilon_{\mu\nu}\partial^{\nu}\omega 
\end{equation}
For definiteness, we consider the Schwinger model (2-dimensional QED), but most of the discussion applies equally well to the 2-dimensional $CP^{N-1}$ sigma model.
As described in the Introduction, we construct a codimension one membrane by including a topological source term
\begin{equation}
\label{eq:source}
S_{\theta} = \frac{1}{2\pi}\int d^2x\; \theta(x) \epsilon_{\mu\nu} F^{\mu\nu} 
\end{equation}
For notational simplicity, we denote the coordinates by $x^1\equiv x, x^2\equiv y$ and take the source field to be a spatial step function at $x=0$, 
\begin{eqnarray}
\label{eq:theta_2D}
\theta(x) & = & \theta_0 \;\;\; x>0 \\
        & = & 0  \;\;\;\; x<0
\end{eqnarray}
Integrating by parts, we see that the source term is equivalent to an ordinary Wilson line operator of the gauge field,
\begin{equation}
\label{eq:Wilsonline}
S_{\theta} = -\theta_0\int A_y dy
\end{equation}
We consider the variation of this term under a U(1) gauge transformation $A_{\mu}\rightarrow A_{\mu}+\partial_{\mu}\omega$. 
If we compactify the $y$ coordinate over a finite range from 0 to $L$,
the variation is given by
\begin{equation}
\label{eq:line-cycle}
\delta S_{\theta} = -\frac{\theta_0}{2\pi}\int_0^L \partial_y \omega\; dy = -\frac{\theta_0}{2\pi}[\omega(L) - \omega(0)] = - \theta_0  n
\end{equation}
Imposing periodic boundary conditions on $g$ requires $n$ to be an integer. Thus the gauge variation of $S_{\theta}$ 
depends on the winding number of the $U(1)$ gauge phase around the compactified $y$-axis.
Gauge invariance of $\exp{iS_{\theta}}$ 
requires that the coefficient $\theta_0/2\pi$ in (\ref{eq:line-cycle}) is an integer.
This is the simplest example of gauge group cohomology, where the 1-cocycle associated with the group element $g=e^{i\omega}$ is just $\omega$, the gauge phase itself.
Since $\delta A_{\mu} = \partial_{\mu}\omega$, the phase (\ref{eq:line-cycle}) is given by the gauge variation of the Chern-Simons 1-form integrated over the Wilson line. 
In the case of 4D Yang-Mills theory, a similar argument applies, where the 1-cocycle is obtained from the descent equations, and is given by the
gauge variation of the Chern-Simons 3-form \cite{Zumino83, Stora83, Faddeev84} integrated over the brane surface.

Now let us allow the brane defined by (\ref{eq:theta_2D}) to fluctuate around its flat starting position at $x=0$. 
The vector $\partial^{\mu}\theta$ is a vector normal to the brane surface and thus specifies its local orientation. In the Schwinger model, this can be identified with the
conserved, non-gauge invariant axial vector current 
$\partial_{\mu}\delta\theta = 2\pi j_5^{\mu}$.
Here $j_5^{\mu}$ is an auxiliary free fermion current which couples to an unphysical massless Goldstone boson in the covariant gauge formulation of the model \cite{Kogut75,Coleman75}.
Its introduction explicitly separates the fermionic component of the axial current from the gauge anomaly.
The gauge invariant current $\hat{j}_5^{\mu}$ is related to the conserved current by
\begin{equation}
\label{eq:gi_current}
\hat{j}_5^{\mu} = j_5^{\mu} + \frac{1}{2\pi}K_{\mu}
\end{equation}
where 
$K_{\mu} = \varepsilon_{\mu\nu}A^{\nu}$. Here,
$\hat{j}_5^{\mu}$ is the physical axial vector current which includes the anomaly.
Note that if we interpret the Wilson line in the usual way as a charged particle world line representing the flow of vector current $j_{\mu}$, 
then $j_5^{\mu} = \varepsilon^{\mu\nu}j_{\nu}$ is always normal to the Wilson line.
The Kogut-Susskind mechanism \cite{Kogut75} has a simple physical interpretation as the separation of a physical charged 
particle into the bare particle and its comoving gauge field. A proper gauge invariant particle state must include both the particle and its surrounding field.
But in order to quantize in a covariant gauge, the KS pole cancellation mechanism must be employed. This introduces two massless scalar fields associated
with the two terms on the right hand side of (\ref{eq:gi_current}). The field representing $j_5^{\mu}$ is an ordinary massless boson field, but the one
representing the gauge anomaly term in (\ref{eq:gi_current}) is a massless ghost field. Physical, gauge invariant amplitudes are constructed with operators
that only contain the gauge invariant sum of the two fields, and massless poles cancel. The physical spectrum has a mass gap given by the 
mass of the Schwinger boson $\mu^2=e^2/\pi$, which is the analog of the $\eta'$ in QCD. The KS mechanism is thus a cancellation between the long range effects which would be induced by separately varying the 
position of a particle and that of its surrounding gauge field. Varying either one separately would induce long range effects, but if
the particle and its surrounding field are varied together, as required physically, the effects are short range, and there are no massless particles
in gauge invariant amplitudes.

When expressed in terms of branes in the gauge field, the Kogut-Susskind mechanism generalizes straightforwardly to the case of 4-dimensional QCD.
In the 2D case the charge carrying object is a pointlike bare fermion, while in 4D QCD it 
is the codimension-one $\theta$ membrane. The Kogut-Susskind mechanism is a cancellation between the massless fluctuations of the membrane surface, described by 
$\partial_{\mu}\theta$, and the wrong-sign massless pole in the Chern-Simons current correlator \cite{TXK}.

\section{Anomaly inflow, transverse brane fuzz, and the Ramond-Ramond field}

\begin{figure}
\vspace*{4.0cm}
\includegraphics{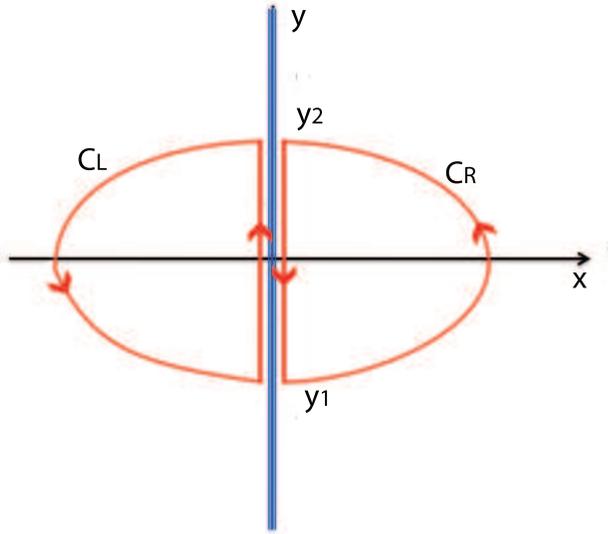}
\vspace{6.5cm}
\caption{Calculating the contribution of a brane along the y-axis to a Wilson loop.}
\label{fig:brane}
\end{figure}

In order to calculate the contribution of a brane to the topological susceptibility in 2D U(1) gauge theory,
consider the calculation of a Wilson loop around a contour ${\cal C}$ that cuts across a membrane, as depicted in Fig. \ref{fig:brane}. 
For simplicity, we first consider the case of a straight brane along the y-axis.
As in the case of a Dirac-Wu-Yang monopole, a description of such a field configuration with no unphysical singularities requires that the $A^{\mu}$ field (as well as the Ramond-Ramond field $\theta$) 
to the left and right of the brane must be written in different gauges $A_L^{\mu}$ and $A_R^{\mu}$. At the
location of the brane along the y-axis, we must match the field description to the left and right of the brane by a gauge transformation $g=e^{i\omega}$ defined on the 
surface of the brane,
\begin{eqnarray}
A_R^y & = & A_L^y-ig^{-1}\partial^y g \equiv A_L^y+\partial^y\omega\;\;\; {\rm for}\;x=0 \\
\theta_R & = &\theta_L+2\pi
\end{eqnarray}

We can now identify the contribution of the membrane to the Wilson loop integral around the contour ${\cal C}$ by writing it in terms of the two closed subcontours ${\cal C}_L$
and ${\cal C}_R$, which do not cross the brane. The Wilson loop integral is given by the sum of the contributions from the two subcontours ${\cal C}_L$ and ${\cal C}_R$, plus a contribution 
from the membrane surface coming from the gauge mismatch between the two contours,
\begin{equation}
\oint_{\cal C}A\cdot dl = \oint_{{\cal C}_L} A_L\cdot dl + \oint_{{\cal C}_R} A_R\cdot dl + \int_{y_1}^{y_2}(A_R-A_L)_y dy
\end{equation}
where $y_1$ and $y_2$ are the two points where the contour ${\cal C}$ punctures the membrane.
The effect of the membrane on the Wilson loop is to add a phase
\begin{equation}
\label{eq:brane_action} 
\int_{y_1}^{y_2} (A_R-A_L)_{\mu}dx^{\mu} = -i\int_{y_1}^{y_2} g^{-1}\partial_{\mu}g\;dx^{\mu} = \omega(y_2) - \omega(y_1)
\end{equation}
This is just the Wu-Yang prescription for the phase of a charged particle propagating in the field of a magnetic monopole: In addition to the 
$A_{\mu}$ phase, if the particle passes from one coordinate patch to another one in a different gauge, the Wilson loop phase receives a contribution
from the gauge transformation that matches the fields along the interface between sections. 

In (\ref{eq:brane_action}) the gauge transformation $g=e^{i\omega}$ which specifies the matching across the brane depends only on the in-brane y-coordinate and appears only in the difference of the 
$A_y$ components of the gauge field. The gauge component $A_x$ transverse to the brane does not enter into the matching.  However, the definition of $g$ as the transformation
which matches the gauge on the two sides of the brane indicates that it should be regarded as localized to the brane at $x=0$. For this reason, we define a gauge transformation
$G\equiv e^{i\Omega}$, where
\begin{equation}
\label{eq:Omega_delta}
\Omega(x,y) = \omega(y)\times \delta(x)
\end{equation}
On a lattice, this would be a gauge transformation that is applied only on a single row of sites along the brane at $x=0$. In the continuum description employed here, the
gauge transformation $G$ always appears in the form of either $\Omega$ or $G^{-1}\partial_{\mu}G= i\partial_{\mu}\Omega$, so (\ref{eq:Omega_delta}) always 
leads to well-defined expressions in terms of delta functions and derivatives thereof.
The separation of a singular bulk gauge transformation at a brane surface into a transformation depending only on the in-brane coordinates multiplied by a delta function in the
transverse coordinate will play a central role in the following discussion.

Small non-topological variations of the gauge transformation $g$ are related by anomaly inflow to local fluctuations of the surface
orientation vector $\partial_{\mu}\theta$. 
In this way, the gauge transformation $g$ which matches the fields on the two sides of the brane 
is promoted to a dynamical field describing fluctuations on the surface of the brane. 
This seems surprising at first, since a gauge transformation should not produce a physical excitation. The key point is that
the transformation of the in-brane component(s) of the $A_{\mu}$ field is not a gauge transformation in the bulk theory unless it
is accompanied by a transformation of the transverse component $A_x$. 
The cohomology of the Wilson line and the brane action (\ref{eq:brane_action}) depends only on the component within the brane $A_y$, and not on the transverse component $A_x$. 
In general, we can distinguish between transformation of the in-brane components of the gauge field, which determines the
gauge cohomology, and transformation of the transverse component, which, by a certain choice of gauge, can be interpreted as
transverse motion of the brane surface.
Let $G\equiv e^{i\Omega}$ be a gauge transformation defined in 2D space-time, $\delta A_{\mu}=-iG^{-1}\partial_{\mu}G$, 
and define separately the transformation of the $x$ and $y$ components of the $A$ field,
\begin{eqnarray}
G_y :\;\;\delta A_x & = & 0 \;,\;\;\delta A_y = -i G^{-1}\partial_y G = \partial_y\Omega\\
G_x :\;\;\delta A_x & = & -iG^{-1}\partial_x G = \partial_x\Omega \;,\;\;\delta A_y = 0 
\end{eqnarray}
where we choose $x$ to be transverse and $y$ parallel to the brane.
For example, a straight, uniform brane along the y-axis at $x=0$ is represented by $G_y$ for the gauge function
\begin{equation}
\label{eq:Omega}
\Omega(x,y) = 2\pi y \times \delta(x)
\end{equation}
giving the transformations
\begin{eqnarray}
G_x :\;\;&\delta A_x  =  2\pi y\delta^{\prime}(x), &\delta A_y  = 0\\
G_y :\;\;&\delta A_x  =  0, &\delta A_y = 2\pi\delta(x)
\end{eqnarray}
While the combined effect of $G_x$ and $G_y$ is a gauge transformation, the transformation $G_y$ by itself inserts a 
physical membrane into the system.
When we generalize this to 4D Yang-Mills, the phase $\omega(y)$ in (\ref{eq:Omega_delta}) will be replaced by a WZW 2-form on the 2+1-dimensional brane.

For the 2D case, we can write any gauge field in the 2D plane in a transverse/longitudinal decomposition,
\begin{equation}
\label{eq:trans-long}
A_{\mu}= \varepsilon_{\mu\nu}\partial^{\nu}\sigma + \partial_{\mu}\Omega  
\end{equation}
A uniform brane along the y-axis corresponds to a gauge configuration
\begin{equation}
\label{eq:inbrane}
A_x=0,\;\;A_y=2\pi\delta(x)
\end{equation}
which is obtained from (\ref{eq:trans-long}) with
\begin{equation}
\label{eq:flatbrane}
\sigma = 2\pi\Theta(x),\;\;\Omega = 0
\end{equation}
(Here we use upper case $\Theta(x)$ to denote a unit step function.) 
The field strength for this configuration is a dipole layer of topological charge,
\begin{equation}
F = 2\pi\delta'(x)
\end{equation}

A general variation of the Chern-Simons current includes both a physical variation $\delta\sigma$ and a gauge variation $\delta\Omega$,
\begin{equation}
\delta K_{\mu} = \varepsilon_{\mu\nu}\partial^{\nu}\delta\Omega - \partial_{\mu}\delta\sigma
\end{equation}
Let's now consider the effect of an infinitesimal transformation applied only to the in-brane component $A_y$ of the brane configuration
(\ref{eq:inbrane}),
\begin{equation}
\label{eq:Avariation}
\delta A_x=0,\;\;\delta A_y=2\pi\varepsilon\delta(x)
\end{equation}
corresponding to the physical variation of the discontinuity of the gauge field across the brane,
\begin{equation}
\label{eq:sigvar}
\delta\sigma = 2\pi\epsilon\Theta(x),\;\;\delta\Omega = 0
\end{equation}
This excitation also varies the density of the topological charge dipole layer,
\begin{equation}
\label{eq:Fvar}
\delta F = 2\pi\epsilon\delta'(x)
\end{equation}
On the other hand, the $A_y$ field in (\ref{eq:Avariation}) can also be obtained by an
in-brane gauge transformation of the form (\ref{eq:Omega_delta}),
\begin{equation}
\label{eq:Omegavar}
\delta\Omega = -2\pi\epsilon y\delta(x)
\end{equation}
This gives the same variation of the in-brane component as in (\ref{eq:Avariation}), 
\begin{equation}
\label{eq:inbrane_transformation}
\delta A_y = 2\pi\epsilon\delta(x)
\end{equation}
but also introduces transverse ``brane fuzz''
\begin{equation}
\delta A_x = 2\pi\epsilon y \delta'(x)
\end{equation}
which cancels the variation (\ref{eq:Fvar}) from $\delta A_y$ to give a net $\delta F=0$.

\begin{figure}
\vspace*{4.0cm}
\includegraphics{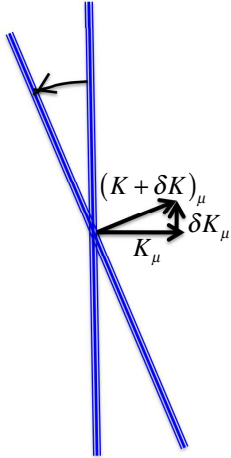}
\vspace{6.5cm}
\caption{For a flat uniform brane, the Chern-Simons current is a vector transverse to the brane surface. Gauge variation of the dual Chern-Simons form on the
brane surface is equivalent to an infinitesimal rotation of the surface orientation.}
\label{fig:orientation}
\end{figure}

We can now combine the variation (\ref{eq:sigvar}) with the gauge transformation (\ref{eq:Omegavar})
to show that the in-brane gauge variation of $A_y$ (\ref{eq:Avariation}) is gauge equivalent to a uniform infinitesimal 
spacetime rotation of the membrane in the $x$-$y$ plane. We perform physical and gauge variations whose combined effect leaves
the $A_y$ component of the field unchanged,
\begin{equation}
\label{eq:sigvar2}
\delta\sigma = 2\pi\epsilon\Theta(x),\;\;\delta\Omega = 2\pi\epsilon y \delta(x)
\end{equation}
This leads to a variation of the Chern-Simons vector that can be interpreted as an infinitesimal spacetime
rotation of the original brane configuration (\ref{eq:inbrane}), as depicted in Fig. (\ref{fig:orientation}),
\begin{eqnarray}
\label{eq:delK1}
\delta K_x & = & 2\pi\epsilon\delta(x) - 2\pi\epsilon\delta(x) = 0 \\
\label{eq:delK2}
\delta K_y & = & -2\pi\epsilon y \delta'(x)\approx 2\pi\left[\delta(x-\epsilon y) - \delta(x)\right]
\end{eqnarray}
This is just the field configuration that would be obtained by an infinitesimal rotation of the original configuration (\ref{eq:inbrane}).
By the anomaly inflow constraint, the gauge variation of the RR field is specified by $\delta(\partial_{\mu}\theta)=-\delta K_{\mu}$,
so (\ref{eq:delK1})-(\ref{eq:delK2}) describes an infinitesimal rotation of the $\theta$ domain wall boundary.

To summarize, if we start with a straight brane along the y-axis, the gauge 
variation of the in-brane component $A_y$, Eq. (\ref{eq:Avariation}) when accompanied by a gauge transformation $\delta\Omega$, Eq. (\ref{eq:Omegavar}), is just an infinitesimal rotation of the brane. 
The variation (\ref{eq:Avariation}) appears as a uniform variation of the topological charge dipole density,
\begin{equation}
\delta F = 2\pi\epsilon \delta'(x)
\end{equation}
Similarly, when we generalize this construction to 4D Yang-Mills theory, the codimension one dipole layers of topological charge (which are in fact
observed in Monte Carlo studies \cite{Horvath03,Ilgenfritz})
represent the brane fuzz associated with the transverse component of the gauge field at the brane surface. 

By repeatedly applying infinitesimal in-brane transformations of the form (\ref{eq:inbrane_transformation}) alternated with bulk gauge transformations, we can 
extend this identification to the case of finite rotations and finite in-brane gauge transformations. 
This provides a novel view of the topological connection between spacetime and the gauge group. In the usual discussion of gauge group topology associated with instantons, one assumes that the
topological charge is localized in spacetime, and that the $A$ field on a circle at infinity is
everywhere gauge equivalent to $A_{\mu} = 0$. The global topology of the gauge field then reduces to the winding number of the
mapping of the spacetime circle to the group phase $e^{i\omega}$, where $A_{\mu}=\partial_{\mu}\omega$. In our discussion of topological charge membranes, topological charge is delocalized, and 
the mapping between the group phase and the spacetime direction arises in a different context. By the anomaly inflow argument, we have related the gauge
phase $\delta \omega$ on the brane to the local orientation angle of the brane surface in the x-y plane. The quantization that results from this mapping 
is not quantization of localized topological charge, but rather, the quantization of the step-function discontinuity of the $\theta$ field across
the brane.

Since the identification between the in-brane gauge transformation and the orientation of the brane surface can be made locally along the brane, the 
previous argument can be extended to describe any infinitesimal fluctuations of the brane. 
As we will discuss in the next Section, the relation imposed by the anomaly inflow constraint between transformation of the in-brane component(s) 
of $A$ and the fluctuations of the brane surface represented by the $\theta$ field 
generalizes to the 4D Yang-Mills case. Combined with the descent equations, this provides a mathematical framework for studying the dynamics of branes
and their role in QCD vacuum structure. 

It is interesting to consider the role of anomaly inflow on the brane in the conservation of axial vector current. As discussed
in Section II, in the presence of quarks the $\theta$ field becomes the $U(1)$ chiral field, and the gauge variation of $\partial^{\mu}\theta$ 
is related to the conserved, gauge noninvariant axial vector current $j_5^{\mu}$. The flow of axial current near the brane is dictated by the anomaly
inflow constraint. Taking the $y$ direction along the brane as Euclidean time, we equate the {\it conserved} axial current $j_{\mu}^5$ to the
variation $\delta(\partial_{\mu}\theta)$ under a gauge transformation of the form (\ref{eq:Omega_delta}), 
\begin{eqnarray}
\label{eq:j5x}
j_5^x & = & \omega'(y)\,\delta(x) \\
\label{eq:j5y}
j_5^y & = & -\omega(y)\,\delta'(x) 
\end{eqnarray}
The flow of spatial current $j_5^x$ into the brane is balanced by the accumulation of chiral charge $j_5^y$ on the brane, giving $\partial_{\mu}j_5^{\mu}=0$.
For example, the gauge function $\omega(y)=\varepsilon y$ describes a constant flow of current into the brane, $j_5^x = \varepsilon\delta(x)$, 
and a linearly increasing chiral charge with time $y$, $j_5^y \equiv j_5^0=-\varepsilon y \delta'(x)$. 

\begin{figure}
\vspace*{4.0cm}
\includegraphics{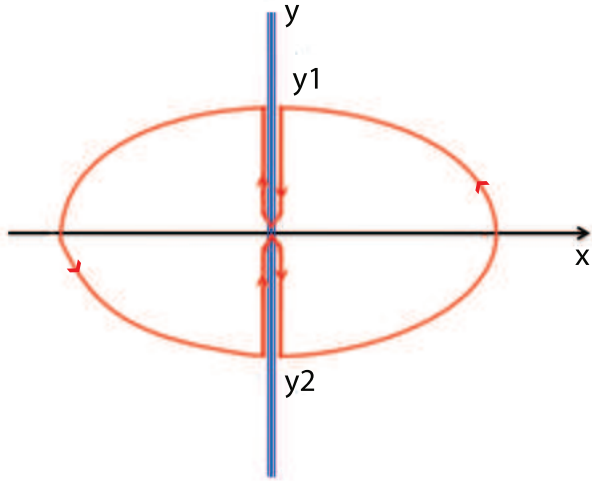}
\vspace{6.5cm}
\caption{Brane puncture induced by the axial anomaly corresponding to quark-antiquark annihilation across the brane.
The Wilson line contribution from the puncture is proportional to the $x$-component of the $A$ field, as in Eq. (\ref{eq:point_splitting})}
\label{fig:brane_puncture}
\end{figure}

The axial anomaly and $\eta'$ mass arise from the possibility of quark-antiquark annihilation between chiral surface modes on opposite sides of the brane. This
causes some of the axial charge to disappear from the brane, leaving a Chern-Simons excitation in the form of a transverse brane fluctuation.
The 2D Schwinger model is an instructive example which suggests the role of transverse brane fluctuations in inducing the quark-antiquark annihilation that gives
the U(1) Goldstone boson a mass. Recall that in that model, the axial $U(1)$ anomaly can be obtained by a simple point splitting method \cite{Kogut75}. We take the y-direction in Fig. 1 to
be Euclidean time, and the axial vector charge $\bar{\psi}\gamma^0\gamma^5\psi = \psi^{\dag}\gamma^5\psi$ may be constructed as a gauge invariant
operator by point splitting,
\begin{equation}
\label{eq:point_splitting}
\hat{j}_y^5= \psi^{\dag}(x+\epsilon)\gamma^5\exp\left[i\int_{-\epsilon}^{\epsilon}A_x dx\right]\psi(x-\epsilon)\rightarrow j_y^5 + A_x
\end{equation}
Here the anomaly $A_x$ arises from the $O(1/\epsilon)$ singularity in the short distance expansion of the quark bilinear.
In Fig. 1 we can interpret the two vertical Wilson lines on opposite sides of the brane as representing a quark bilinear 
which straddles the brane. The point splitting procedure (\ref{eq:point_splitting}) suggests that the matching at
the brane surface, Eq. (\ref{eq:brane_action}), is sensitive to not only the in-brane gauge field component $A_y$, but also to the transverse $A_x$ component.
The form of the line integral (\ref{eq:brane_action}) indicates that such an effect would arise from a fluctuating brane whose world line deviates
from the y-axis, picking up a contribution from the x-component of the gauge field. In the picture where the Wilson lines in Fig. 1
are the fermions (quarks) of the Schwinger model, the brane fluctuation term induces quark-antiquark annihilation into a pure gauge excitation
of the Chern-Simons tensor as depicted in Fig. \ref{fig:brane_puncture}. As discussed in Ref. \cite{TXK}, this is the origin of the 4-quark contact term that
is responsible for the $\eta'$ mass.  

Note that, although the 2D gauge function $\Omega(x,y)$ in (\ref{eq:Omega}) is dimensionless, we are implicitly taking $x$ and $y$ to be given in units of the
physical length scale. For example, on the lattice, the observed membranes are of finite thickness in lattice units, but become singular delta-functions
in the continuum limit. 
In the 2D Schwinger model, the length scale is determined by the gauge coupling constant $e$ which has dimensions of mass,
while in the $CP^{N-1}$ models and in QCD, it is determined dynamically in terms of the cutoff scale by asymptotic freedom.
In QCD, the relevant length scale is set by the pseudoscalar decay constant $\ell\sim f_{\pi}^{-1}$.

\section{Membranes in 4D Yang-Mills Theory}

As in the 2D example, the anomaly inflow constraint allows one to reduce the gauge dynamics of 4D Yang-Mills theory at the codimension one surface of 
a brane to a lower-dimensional theory on the brane surface coupled to a bulk $\theta$ field. 
In the 2D U(1) case, the matching of gauge fields across the brane depicted in Fig. 1 gives a contribution to the Wilson loop phase proportional to the length of the membrane,
\begin{equation}
\int_{y_1}^{y_2}\delta A\,dy=\omega(y_2)-\omega(y_1) = 2\pi(y_2-y_1)
\end{equation}
We interpret this as the action associated with the membrane world line between $y_1$ and $y_2$.
We may think of the 1-cocycle $\omega(y)$ as a phase attached to the pointlike brane at a fixed time $y$.
In 4D Yang-Mills theory, the world volume action of the brane is the gauge variation of the 3D Chern-Simons tensor, and
the 1-cocycle obtained from the descent equations (\ref{eq:CSvariation})-(\ref{eq:K_2B}) plays the role of the
Hamiltonian density for the 2-dimensional brane at a fixed time. The gauge transformation $g$ becomes a local field on the world sheet of the brane describing its fluctuations
in the bulk space.

The approach we use for constructing a topological charge membrane in 4D Yang-Mills theory is the same as in 2D U(1). We add a brane which spans three of the four Euclidean dimensions
by including a theta term which is a step function in the transverse coordinate $x_4\equiv x$ and independent of the other coordinates, as in Eq. (\ref{eq:theta}). Again, integrating by parts,
we write the action $S_{\theta}$ as an integral localized to the brane surface,
\begin{equation}
\label{eq:Wbag}
 S_{\theta} = -\int d^4x \partial^{\mu}\theta \; K_{\mu} = -\theta_0\int_{R_3} {\cal K}_3^{\alpha\beta\gamma}dx_{\alpha}\wedge dx_{\beta} \wedge dx_{\gamma}
\end{equation}
where the Chern-Simons current $K_{\mu}$ and the dual CS tensor ${\cal K}_3^{\alpha\beta\gamma}$ are given by (\ref{eq:CScurrent}) and the last integral is over the 3-dimensional brane world volume.. 
The integral of the 3-index CS tensor $ K_3$ over a 3-dimensional surface has been referred to by Luscher \cite{Luscher78} 
as a ``Wilson bag'' operator. In the discussion of topological charge structure of 4D Yang-Mills theory, it plays a role analagous to the Wilson line in the 2D U(1) case.
Just as the Wilson line can be interpreted as the gauge phase attached to the world line of a charged particle, the Wilson bag integral is the gauge phase associated
with the world volume of a 2+1 dimensional membrane. Note also that the value of a closed Wilson loop in 2D U(1) theory is equal to the total topological charge 
contained inside the loop. Similarly, the integral of ${\cal K}_3$ over a closed Wilson bag is equal to the amount of Yang-Mills topological charge contained in the
4-volume enclosed by the bag.

If we keep the brane flat by holding the discontinuity of $\theta(x)$ fixed at $x=0$, the action $S_{\theta}$ is not gauge invariant.
The gauge variation of $K_{\mu}$ is given by the sum of two terms, Eqns. (\ref{eq:K_2A}) and (\ref{eq:K_2B}). The term (\ref{eq:K_2A}) depends only on $g$,
and is proportional to the winding number density,
\begin{eqnarray} 
\label{eq:winding}
w(x) & = & \frac{1}{24 \pi^2}\epsilon_{\alpha\beta\gamma}{\rm Tr}\left[g^{-1}\partial^{\alpha}g\;g^{-1}\partial^{\beta}g\;g^{-1}\partial^{\gamma}g\right]\\
     & = &  \frac{1}{8 \pi^2} \epsilon_{\alpha\beta\gamma}\partial^{\alpha}K_{2A}^{\beta\gamma}
\end{eqnarray}
where ${\cal }K_{2A}^{\alpha\beta}$ is the WZW term (\ref{eq:K_2A}). The mod $2\pi$ ambiguity of the surface integral of ${\cal K}_{2A}^{\alpha\beta}$ again leads to the 
requirement that the coefficient $\theta_0/2\pi$ be integer quantized to maintain gauge invariance of the exponentiated WZW term.

Invoking the anomaly inflow constraint, we again require that $\partial_{\mu}\theta$ transform under a Yang-Mills gauge
transformation $g=e^{i\omega}$, so that it cancels the variation of the Chern-Simons current given by (\ref{eq:CSvariation}),
\begin{equation}
\label{eq:inflow}
\delta(\partial_{\mu}\theta) = - \delta K_{\mu} = \epsilon_{\mu\alpha\beta\gamma}\partial^{\alpha}\left({\cal K}_{2A}^{\beta\gamma} + {\cal K}_{2B}^{\beta\gamma}\right)
\end{equation}
The WZW term ${\cal K}_{2A}^{\beta\gamma}$, Eq. (\ref{eq:K_2A}) is analogous to the gauge phase $\omega$ in (\ref{eq:Omega_delta}) for 2D U(1). As in that case, a gauge 
transformation on $K_{\mu}$ induces a fluctuation of the vector $\partial_{\mu}\theta$, which represents a fluctuation of the brane surface.
To study this further, we consider a truncated brane with a finite boundary by taking the source field $\theta(x)$ to be a constant inside a 
3-dimensional ball of radius $R$ carved out of the Euclidean brane, representing the propagation of a 2-dimensional disk over a finite time interval:
\begin{eqnarray}
\theta(x) &  = & \theta_0 \;\;\; x_1>0,\; \sqrt{x_2^2 + x_3^2+x_4^2} < R \\
          & = & 0 \;\;\; {\rm otherwise}
\end{eqnarray}
Then the gauge variation of the action $S_{\theta}$ can be written as a 2-dimensional action on the surface of the ball,
\begin{equation}
\label{eq:delta_S}
\delta S_{\theta} = \int_{S_2} dx_{\alpha}\wedge dx_{\beta}\left[\frac{1}{3}{\rm Tr}\left(\omega\; g^{-1}\partial^{\alpha}g\;g^{-1}\partial^{\beta}g\right)
+ {\rm Tr}\left(\partial^{\alpha}g\;g^{-1}A^{\beta}\right)\right] + {\cal O}(\omega^4) 
\end{equation}
Note that $\delta S_{\theta} = S_{\theta}(g) - S_{\theta}(1)$, so that the expression (\ref{eq:delta_S}) is the $g$-dependent part of the action..
For a given Yang-Mills potential $A^{\gamma}$ the equation (\ref{eq:inflow}) allows us to translate a color gauge transformation $g$ into a fluctuation of the
brane orientation vector $\partial_{\mu}\theta$. Thus the first term in the action (\ref{eq:delta_S}) describes the self interaction of the fluctuations
of the brane inside the 3-volume of the ball in terms of a 2D WZW model on the surface of the ball. 
In a Hamiltonian framework, the 3D ball represents at fixed time a 2-dimensional spatial disk of maximum radius $R$, with the Kac-Moody currents of the 
WZW model flowing around the boundary of the disk. 

Just as we did in the 2D case, we may study the contribution to topological susceptibility of a Yang-Mills membrane in the vacuum by determining its effect
on a probe Wilson bag operator that is cut into two sections by the 3D plane of the membrane. As before, the gauge choice on the two sides of the membrane
must differ by a relative gauge transformation $g$. The analog of the Wu-Yang phase (\ref{eq:brane_action}) is the 1-cocycle of the gauge transformation $g$, 
given by Eq. (\ref{eq:delta_S}). As in the 2D case, a straight flat brane can be introduced by a gauge transformation which transforms the 3 in-brane components of the gauge field
by a topologically nontrivial gauge transformation. Of the four Euclidean coordinates $x_{\mu},\; \mu=1,\ldots 4$, we denote the 3 coordinates within the brane by $x_i \equiv y_i, i=1,\ldots 3$,
and the transverse coordinate by $x_4 \equiv x$. For simplicity we will discuss $SU(2)$ gauge theory, but generalization to $N_c>2$ is straightforward.
To construct a brane at $x=0$ we perform a gauge transformation on the 3 in-brane components of the Yang-Mills field by an $SU(2)$ phase
\begin{equation}
\label{eq:omega_inbrane}
\omega = -i\log\;g = \pi \frac{\vec{y}\cdot\vec{\sigma}}{\ell}
\end{equation}
The topological WZW term of the corresponding 1-cocycle, Eq. (\ref{eq:K_2A}), has the form
\begin{equation}
{\cal K}_{2A}^{\beta\gamma} = \frac{\pi}{3}\epsilon^{\alpha\beta\gamma}y_{\alpha}
\end{equation}
This is embedded in 4-dimensional space by restricting it to the 3D surface of the brane at $x=0$ with a delta-function.
Then the gauge variation of the Chern-Simons current is given by
\begin{equation}
\delta K_{\mu} = \frac{\pi}{3}\epsilon_{\mu\alpha\beta\gamma}\partial^{\alpha}\left({\cal K}_2^{\beta\gamma}\times \delta(x)\right)
\end{equation}
The quantity $\partial^{\alpha}{\cal K}_2^{\beta\gamma}$ consists of the two terms in the gauge variation of the Chern-Simons 3-form on the membrane surface, Eq. (\ref{eq:CSvariation2}). 
For the topological term ${\cal K}_{2A}$, this gives
\begin{equation}
\label{eq:deltaK}
\delta K_{\mu} = \frac{\pi}{3}\epsilon_{\mu\alpha\beta\gamma}\partial^{\alpha}\left(\epsilon^{\beta\gamma i}y_i \delta(x)\right)
\end{equation}
Anomaly inflow for the CS current $K_x$ transverse to the brane shows that the in-brane gauge transformation (\ref{eq:omega_inbrane}) creates a uniform
codimension one brane transverse to the x-axis. From Eq. (\ref{eq:deltaK}) we find
\begin{equation}
\partial_x\theta = -\delta K_x = - 2\pi\delta(x)
\end{equation}
Once again, as in the 2D discussion, we consider an additional infinitesimal in-brane transformation of the same form,
\begin{equation}
\label{eq:domega_inbrane}
\delta\omega = \pi \epsilon\frac{\vec{y}\cdot\vec{\sigma}}{\ell}
\end{equation}
Transforming only the in-brane components, this varies the $\theta$ discontinuity across the brane
\begin{eqnarray}
\delta(\partial_x\theta) & = & -\delta K_x = - 2\pi\epsilon\delta(x) \\
\delta(\partial_{y_i}\theta) & = 0 & 
\end{eqnarray}
Following the same argument applied to the 2D Wilson line excitation, we can apply a 4D Yang-Mills gauge transformation to write this in a form where
the discontinuity of $\theta$ across the brane remains $2\pi$ (to first order in $\epsilon$), but the local orientation of the brane surface has rotated slightly,
\begin{eqnarray}
\delta(\partial_x\theta) & = & 0 \\
\delta(\partial_{y_i}\theta) & = & 2\pi\left[\delta(x+\varepsilon y_i) - \delta(x)\right] \approx 2\pi\varepsilon y_i \delta'(x)
\end{eqnarray}

We have shown that a small fluctuation of the 3-dimensional in-brane gauge transformation $g=e^{i\omega}$ is equivalent to a fluctuation of the surface in
the transverse space. Note that $g$ is defined entirely on the 3-dimensional brane without any reference to the transverse coordinate. The intepretation of $g$ as
describing a transverse fluctuation of the brane arises when we embed the 3-dimensional gauge transformation $g$ in the 4D gauge configuration with a transverse
delta function. The relation between gauge variations on the brane and transverse fluctuations is reminiscent of similar connections in string theory. 
In the case of gauge theory, this connection is a direct consequence of the descent equations and cohomology structure of Yang-Mills theory,
which describes the interplay between gauge variations and spacetime derivatives. The gauge variation of the Chern-Simons 3-form as a functional of $g$ 
plays the role of a world sheet action for the brane. The descent equations express this 3-dimensional action as the exterior derivative of a WZW 2-form ${\cal K}_2^{\alpha\beta}$,
Eq. (\ref{eq:K_2A}) and (\ref{eq:K_2B}). 

For the 2-dimensional U(1) case, the brane action depends only on the gauge phase $\omega$. The analogous term in 4D Yang-Mills is the topological WZW term, ${\cal K}_{2A}$
integrated over the spatial surface of the brane.
This also depends only on the gauge transformation, but unlike the 2-dimensional case, the WZW action includes self-interactions for the membrane fluctuations. Another
new feature of the 4D Yang-Mills case is the additional, nontopological term in the action, ${\cal K}_{2B}$, which defines a coupling between the color Kac-Moody current
associated with the WZW field $g$ and the color gauge field $A_{\mu}$. This term describes the emission of a gluon from a fluctuating brane. 

\section{Discussion}

Large $N_c$ chiral lagrangian arguments, gauge-string holography, and Monte Carlo results all indicate that the topological structure
of the QCD vacuum is dominated by codimension one membranes which appear as dipole layers of topological charge, i.e. juxtaposed positively and negatively charge sheets.
In this paper we have discussed an approach to the dynamics of these membranes based on their interpretation as Wilson bags \cite{Luscher78},
i.e. singular, sheet-like excitations of the Chern-Simons tensor on codimension one surfaces. A Wilson bag plays the role of a domain wall between
local quasi-vacua with discrete values of the QCD $\theta$ parameter differing by $\pm 2\pi$. Holographically, the Wilson bag is interpreted as a D6 brane
in IIA string theory, carrying Ramond-Ramond charge.  The analogy with Wilson line excitations in 2-dimensional U(1) gauge theory is very instructive.
In Coleman's original discovery of the topological $\theta$ parameter in the massive Schwinger model \cite{Coleman75}, he showed that 
$\theta$ could be interpreted as a background electric field. A domain wall between vacua with different values of $\theta$ is just a charged particle world line. The associated Wilson line
integral of the $A$ field can be reinterpreted as a surface integral of the Chern-Simons flux, which is the 2D analog of a Wilson bag.
As discussed in \cite{TXK}, the Ramond-Ramond field $\theta$ plays the role of the background electric field in the 4D Yang-Mills generalization of Coleman's discussion.

The approach we have pursued in this paper avoids any direct use of the holographic framework to introduce branes into QCD. Instead, a membrane is constructed from its 
4-dimensional definition as a discrete step in the QCD $\theta$ parameter, or equivalently, a surface integral of the Chern-Simons tensor. This approach allows us 
to address questions of brane dynamics in the powerful mathematical framework of gauge group cohomology, anomaly inflow, and the descent equations of Yang-Mills theory \cite{Stora83, Zumino83, Faddeev84, Callan-Harvey}.
The anomaly inflow constraint at the brane surface defines the connection between the $\theta$ field and the gauge field. It can be thought of as ``Gauss's law'' for the
$\theta$ field, with the source term given by the Chern-Simons tensor on the brane. This implies a nontrivial transformation of the $\theta$ field under
Yang-Mills gauge transformations. This transformation is specified by the gauge variation of the Chern-Simons tensor, as expressed by the
descent equations of Yang-Mills theory \cite{Faddeev84}. The sequence of arguments is 
simplest in the 2-dimensional $U(1)$ case, where the relevant descent equation is just the gauge transformation itself $\delta A_{\mu}=\partial_{\mu}\omega$.
In 4D Yang-Mills, the analog of $A_{\mu}$ is the 3-index Chern-Simons tensor, and its gauge variation defines a 1-cocycle $\omega^{\mu\nu}$, which is a Wess-Zumino-Witten
2-form $K_2^{\mu\nu}$ integrated over the 2-dimensional spatial surface of the brane. It is a functional of the gauge group variation $g$ on the brane which
appears as the WZW field. In the same sense that the gauge phase $\omega$ along the timelike Wilson
line in 2D can be thought of as the phase attached to the wave function of a pointlike charged particle, the 1-cocycle $\omega^{\mu\nu}$ can be thought of as the gauge phase attached
to a 2-dimensional membrane in the $\mu$-$\nu$ plane.

The results presented here suggest a very appealing model for the chiral condensate.
In the membrane vacuum, the near-zero Dirac eigenmodes which are needed to form a condensate
appear as surface modes on the topological charge membranes. The fact that the membrane consists of opposite-sign topological charge sheets on opposite sides
of the brane implies that left- and right-handed chiral densities $\overline{q}(1\pm\gamma_5)q$ will appear on opposite sides of the same brane. We saw in Section III that
conservation of axial vector current near the brane surface arises from a balance between the current impinging on the brane from the transverse direction and
the current flowing along the brane. The axial $U(1)$ anomaly arises by the following physical mechanism: when a membrane fluctuates the quark and antiquark
states on opposite sides of the brane will overlap and thus can annihilate, as in Fig. \ref{fig:brane_puncture}, if the quark and antiquark are of the same flavor. This is the origin of the $\eta'$ mass insertion and the nonconservation 
of axial $U(1)$ current. If we suppress the $\bar{q}q$ annihilation process (either by taking the large $N_c$ limit, or by introducing two flavors of quark and
considering flavor nonsinglet pions), this picture also provides an understanding of massless Goldstone boson propagation. 

It was argued in Ref. \cite{TXK} that
the Ramond-Ramond field in QCD gives rise to effective 4-quark contact terms responsible for both the $\eta'$ mass insertion and a Nambu-Jona Lasinio-type interaction 
that provides the attractive interaction between chiral pairs that produces the $\bar{q}q$ condensate. 
To see the connection between the Ramond-Ramond field and Goldstone bosons, we recall the equivalence between a 
rotation of the QCD $\theta$ parameter and a variation of the flavor singlet
chiral field $\eta'$. In the usual discussion this equivalence follows from the index theorem. Our discussion exhibits a physical mechanism for this connection
by identifying the quark near-zero modes as surface modes of the topological charge membranes. 
The anomaly inflow formalism defines a spacetime dependent $\theta(x)$ which is sourced by
singular Chern-Simons excitations of the gauge field. The discontinuities of the $\theta$ field define the location of the membranes. The connection between $\theta$ and the
chiral field follows from the assumption that the condensate lives on the brane surfaces. This leads to the identification of $\partial_{\mu}\theta$ as
the axial vector current. In a vacuum filled with a ``topological sandwich'' of membranes \cite{Ahmad05, Thacker06, Thacker10}, long wavelength Goldstone bosons propagate 
masslessly via chiral quark pairs occupying delocalized surface modes on the branes combined with a collective transverse oscillation of the branes. 
The bulk oscillation and surface mode propagation are locked together by 4D gauge invariance and the anomaly inflow constraint, which balances the bulk and surface
currents to give massless Goldstone boson propagation.  

\section*{Acknowledgments}

This work was supported by the Department of Energy under grant DE-FG02-97ER41027. CX is supported by the research funds from the Institute of Advanced Studies, Nanyang Technological University, Singapore.

\end {document}